\documentclass[runningheads]{llncs}
\usepackage{graphicx}
\usepackage{float}
\usepackage{subfigure}
\usepackage{hyperref,xcolor}
\usepackage[english]{babel}
\usepackage{enumitem}

\begin{document}
\title{Integrating Dark Pattern Taxonomies}

 \author{
     Frank B.W. Lewis\inst{1}\orcidID{0000-0001-6929-4455} \email{fbl773@usask.ca}
     \and Julita Vassileva\inst{1}\orcidID{0000-0001-5050-3106}\email{jiv@cs.usask.ca}
 }
 \institute{University of Saskatchewan, Saskatoon SK S7N 5C9, Canada}
 \authorrunning{F. Lewis et al.}

\maketitle

\begin{abstract}
The problem of ``Dark Patterns" in user interface/user experience (UI/UX) design 
has proven a difficult issue to tackle. Malicious and explotitative design has 
expanded to multiple domains in the past 10 years and which has in turn led to 
multiple taxonomies attempting to describe them. While these taxonomies holds 
their own merit, and constitute unique contributions to the literature, their 
usefulness as separate entities is limited. We believe that in order to make 
meaningful progress in regulating malicious interface design, we must first 
form a globally harmonized system (GHS) for the classification and labeling of 
Dark Patterns. By leaning on network analysis tools and methods, this paper 
synthesizes existing taxonomies and their elements through as a directed graph. 
In doing so, the interconnectedness of Dark patterns can be more clearly revealed 
via community (cluster) detection. Ultimately, we hope that this work can serve as the 
inspiration for the creation of a glyph-based GHS for the classification of 
Dark Patterns.

\end{abstract}

\section{Introduction}\label{P1}

After various catastrophes regarding the inadequate research and risk
communication regarding chemicals, it was decided at the 1992 Rio Earth summit 
to begin the formation of a Globally Harmonized System (GHS) for the 
classification and labeling of chemicals\cite{GHS}. An international team of 
experts was posed with the task of creating immediately intelligible warning 
labels and data-sheets that could unify hazard communication globally. This 
was a response to tragedies like DDT, Asbestos, and Thalidomide. These 
products did \textit{exactly what they promised}. DDT kills insects, 
asbestos is an excellent insulator, thalidomide relieves symptoms of morning 
sickness. However, the unintended damage caused by their use was immense. 
Similarly, the design of shopping apps help in finding something to buy, 
the design of social media applications  facilitate connection, and video 
games can be a pleasurable experience that builds communities and forms 
friendships. Unfortunately, often cooperate greed is the driving force behind 
system design, and the so called "Dark Patterns" \cite{1_Brignull_2010} in the 
interface design transform seemingly harmless benefit-providing systems into 
instruments of user exploitation. Predatory Monetization\cite{14_Petrov_2021}, 
exploiting obsession\cite{0_Ahuja_2022}, and disguised data collection\cite{4_Greenberg_2014} 
are examples of dark patterns in the design of interactive technologies. 
The massive and unconscious exposure to dark patterns can lead users to waste 
time and money, to frustration, and even to mental health issues. Yet many users 
are either unaware that they are victims of deliberate design "traps" or 
acquiesce to use the application because it offers some functionality they need. 
If there were a unified system for labeling dark patterns in user interface 
design, it would be possible to regulate their use similar to the regulation of 
harmful chemicals. It would also be possible to educate consumers and create 
appropriate warnings in the description of interactive applications that use 
such patterns. Our vision is that a Globally Harmonized System for Dark Patterns 
in software interface design is needed and our work presented in this paper is 
a step in this direction.

\section{Background and Related Work}\label{P2}

This section provides an overview of the previous works on dark patterns that we are aware of and defines the research questions we set out to answer. 

\subsection{Regarding ``Dark" Patterns}
The term "Dark Pattern" (DP) was coined by Harry Brignull, a UK-based user
experience designer and consultant, in 2010. Brignull identified deceptive User Interface (UI) techniques that trick users into actions they might 
not have otherwise taken and created a website to raise awareness about 
them. This term has since become widely recognized in discussing and 
critiquing unethical design practices in digital interfaces. Recently, 
there has been some controversy over the phrase, and motions have 
been made to replace the term with ``Deceptive"/``Manipulative" Design. We 
consciously made the choice to stick with the original term because  
alternatives (i.e. ``Deceptive") describe only a part of what 
such patterns can achieve. Manipulation, coercion, and exploitation 
are equally common outcomes of interaction with these designs. Thus
relegating them to simple deception limits our perception of 
their depth. This is not to downplay the legitimate concern 
surrounding the use of the term ``Dark" patterns, related to a 
possible connotation with race, but for a lack of a commonly used 
and familiar term in UI and UX literature that encompasses all 
the possible different aspects of these designs, we will use ``Dark" 
in this paper. Next, we present briefly the existing classifications 
of dark patterns since 2010.

\subsection{Related Works}
The first taxonomy of dark patterns by \textbf{Brignull (2010)} \cite{1_Brignull_2010} was derived through personal observations and 
reporting, without a documented methodology of its creation. Although Brignull's work lacks 
scientific rigour, it captures the essence of dark patterns and is considered 
a seminal work in the area. 
The taxonomy of \textbf{Conti \& Sobiesk (2010)} \cite{2_Conti_2010} focused on dark patterns across domains. 
A year-long study involving  22 undergraduate students gathering dark patterns 
was combined with input from roughly 75 participants of the ``Hackers of Planet Earth" 
Conference to create a multi-dimensional taxonomy of Dark Patterns.
\textbf{Zagal et al. (2013)} \cite{3_Zagal_2013} focused on the 
domain of games. It was constructed by considering system interaction as a 
contract from which ethical boundaries could be derived between users and 
designers.
\textbf{Greenberg et al. (2014)} \cite{4_Greenberg_2014}, using the ``affinity 
diagramming" practice, clustered and speculated upon examples of potential dark-pattern utilization within the context 
of Proxemic Interactions .
Rooted in the domain of user privacy, 
\textbf{Bosch et al. (2016)} \cite{5_Bosch_2016} constructed their taxonomy by considering established privacy 
design patterns and deriving dark patterns from their antithesis. These derived patterns 
were then evaluated against the known DPs at the time (i.e. Brignull's). This 
work was considered as a potential basis seeing as its roots in design standards 
makke it domain independent, however its focus on privacy makes it less generally 
applicable than Ahuja et. al's\cite{0_Ahuja_2022}.
\cite{1_Brignull_2010}).
\textbf{Grey et al. (2018)}'s taxonomy \cite{6_Gray_2018} was constructed 
from examples across domains. Examples of dark patterns were collected, catalogued, and 
sourced iterativly over 2 months. The constant comparison method\cite{GT_Glaser_Strauss} was then 
applied along with document analysis to describe the nature of the dark patterns 
found, categorized according to Brignull's\cite{1_Brignull_2010} taxonomy before 
being re-categorized via open coding. The resulting taxonomy is thus structured 
around dark strategies and designer motivation rather than around context and content 
as Brignull's\cite{1_Brignull_2010}.
The \textbf{Norwegian Consumer Council (NCC) (2018)} analyzed the 
design and wording of update prompts from  Facebook, Google, and Microsoft from 
a data protection standpoint after GDPR regulations were enacted\cite{7_NCC_2018}.
Situated in the domain of ``Home 
Robots", \textbf{Lacey \& Caudwell (2019)} \cite{8_Caudwell_2019} did not explicitly propose a taxonomy, but their work was flagged as such by 
Ahuja et al.\cite{0_Ahuja_2022} for its proposal that ``Cuteness" can be a dark pattern 
because a robot's ``cuteness" encourages users to overlook or 
completely miss their devices' data-collecting capabilities.
By crawling and scraping image samples 
from various e-commerce sites, \textbf{Mathur et al. (2019)} \cite{9_Mathur_2019} abstracted a 
set of five dark pattern \textit{attributes}, which were then applied to 15 
dark patterns within 7 super-categories, some of which have 
direct counterparts in Brignull's  taxonomy \cite{1_Brignull_2010}.  The 
\textbf{National Commission for Informatics and Liberty (CNIL) (2020)} \cite{10_CNIL_2020} built their taxonomy 
with the goal of consumer protection, though no formal methodology was documented in the document. Many of its items tie directly to Brignull's original set (i.e. ``Trick Question"\cite{1_Brignull_2010,10_CNIL_2020})
while others simply go by another name (i.e. Brignull's ``Bait and Switch"\cite{1_Brignull_2010} 
and CNIL's ``Bait and Change"\cite{10_CNIL_2020}). 
\textbf{Grey et. al (2020)}'s work \cite{11_Gray_2020} focused on 
``Asshole Design" as flagged by users of the Social Media platform Reddit.
Posts on Reddit were analyzed and open coded through several rounds of 
iterative content analysis. This process rendered six dark patterns.
\textbf{Bongard-Blanchy et al. (2021)} administered an online survey to 
participants with the aim of gauging user awareness of dark patterns. The patterns 
used to measure user awareness constitute Bongard-Blancey's 
contribution to the area of dark pattern taxonomies\cite{12_Bongard_2021}.
Using techniques inspired by Value Sensitive 
Design and scenario construction, \textbf{Mhaidli \& Schaub (2021)} \cite{13_Mhaidli_2021} critically examined some potential advances in 
virtual/augmented/mixed reality  (XR) technology 
regarding their ethical implications. Scenarios and narratives were constructed 
and used to generate a set of five dark patterns that could operate in this 
space.
\textbf{Petrovskaya \& Zendle (2021)} \cite{14_Petrov_2021} focus on the issue of Predatory Monetization in the domain 
of games. By using the United 
Kingdom's unfair trading regulations, researchers released a survey requesting 
reports of unfair/aggressive monetary encounters online. The results were 
transformed into a two-tiered collection of 36 dark patterns in 
eight super categories.
\textbf{Westin \& Chiasson (2021)} \cite{15_Westin_2021} likewise focused on a
single dark pattern - ``Fear of Missing Out (FoMO) centric design" - in the context 
of user privacy. By utilizing inductive grounded theory, researchers 
dimensionalized and categorized design interactions to pose the idea that the 
core issue is not with individual patterns but with their aggregated effects 
within ``dark digital infrastructure" .
\textbf{Wu et al. (2021)} generated their taxonomy by 
observing and coding live-stream shopping interactions on Taobao and TikTok. 
This taxonomy is interesting, but focuses more on the habits of the \textit{streamers}
than on interface design\cite{16_Wu_2021}. Many of their identified patterns have been removed 
from our integration process for this reason.
The work of \textbf{Ahuja et al. (2022)}
unifies existing taxonomies with respect to their relation to user agency. This is, to our best knowledge, the most recent work in the area.  It was able to describe and categorize many existing taxonomies and provides a good basis for further unification and integration. 
  
\subsection{Summary and Outlook: Synthesizing Existing Taxonomies}
As previous taxonomies become the fuel for the next, existing literature 
seems to have entered an endless cycle of splitting and 
joining dark pattern types. 
It appears that research in Dark Pattern taxonomies has reached a saturation point and researchers should
begin the work of their integration to create a global standard if we are to 
mitigate dark patterns to any meaningful capacity - a process which this work strives to  begin. 

Our work represents an effort in this direction. We 
do not define any \textit{new} dark patterns but instead seek to reveal which of 
the \textit{existing} ones are prevalent, which ones are so similar that can be considered the same,  and to uncover how dark patterns relate to each 
other by representing all taxonomies within a single graph (with a couple of exceptions that are too domain specific). Using methods
borrowed from network analysis we reveal naturally forming communities of dark patterns. We must clarify that we use the term "community" in the sense  it is used in the area of network analysis, as a group of nodes that are similar in some specific features.  The dark pattern communities provide a level of abstraction that can simplify the task of identifying the pattern employed by a suspect UI as it can be compared / contrasted with other patterns in the relevant community. 
Our future work will define also heuristics against which suspect UI designs can be evaluated with respect to their "darkness".

The paper focuses on the following research questions:
\subsection{Research Questions}\label{RQ}
\begin{itemize}
	\item{\textbf{RQ1}: What general Dark Pattern communities emerge from existing literature?}
	\item{\textbf{RQ2}: Can a process of heuristic evaluation be used to identify Dark Patterns based on these emerging communities?}
\end{itemize}

In the next section we present the methodology we used to construct the integrated Dark Patterns taxonomy and the resulting communities.

\section{Methodology}

To construct this integrated taxonomy, we started with the 
most comprehensive taxonomy to date (Ahuja et al. 2022\cite{0_Ahuja_2022}), 
and reviewed the papers relating to its contained taxonomies. In fact, the early versions of our taxonomy 
were simply verbatim graphical representations of Ahuja et al.'s work. What we 
add to Ahuja et al.'s, as well as other taxonomies, is that we make explicit the relationships between 
their contained dark patterns. This allows to describe in detail exactly how they relate 
within different contexts.

When seeking to graphically represent an integrated taxonomy, many of the previously 
mentioned taxonomies were considered as a base-taxonomy. 
We settled on using Ahuja et. al's work they have unified more previously defined
taxonomies similar works. What further distinguishes this taxonomy is its roots 
in philosophical conceptualizations of autonomy. By basing the taxonomy in ideals 
rather than specific domains, applications, or common industry practice, its 
potential for generalization exceeds that of many taxonomies while being able to 
completely describe them. Thus, it provides a solid foundation for building 
systems that can track and identify recurring dark patterns across domains. 
Moreover, while Ahuja et al.'s work does define some of its own taxonomy items, 
the majority of them come from previous work. In doing so, Ahuja et al. not 
only provides a common mapping, but does not further dilute the Dark Pattern 
space, making the construction of our initial graph trivial.

\subsection{Process} \label{sec:graph_construction}
Our first step was to represent he Ahuja et al.'s,  taxonomy 
graphically. We chose to use a directed graph with two node types and two edge 
types. \textit{Taxonomy Nodes} are nodes with represent taxonomies, and are 
named after their first author. These nodes have only outward edges toward 
\textit{Pattern Nodes}. Pattern nodes have incoming edges from either patterns 
that \textit{implement/employ them}, or from the taxonomy they belong to.
Likewise, pattern nodes only have outgoing edges to patterns that they 
implemented, or employed (utilized).  This allows us to reveal which patterns are most 
commonly used/implemented across the taxonomies by finding the pattern node
with the highest number of incoming links or \textit{in-degree}. 

In the first version of the graph, Ahuja et al.'s taxonomy was followed exactly 
with no modifications (i.e. merging similar patterns). Minor versions 
(i.e. 2.1, 3.2) were formed by analyzing the literature surrounding similar 
patterns, and making an informed decision to merge them into a single node. 
Each iteration of this process constituted a minor revision. A log of 
which nodes were merged and why can be found in our project
\href{REMOVED_FOR_ANONIMITY}{wiki}(LINK REMOVED FOR BLIND REVIEW).

We used an iterative process of community detection and node merging based on 
pattern similarity to refine our graph. For this we used a software tool for 
network analysis called \href{https://gephi.org/}{Gephi}. 

\textbf{Taxonomy Version 1:}
The first version of the taxonomy (1.0) is a verbatim representation of Ahuja et 
al.'s\cite{0_Ahuja_2022} original taxonomy node/edge policy. Once this 
taxonomy had been added, each of the taxonomies surveyed by Ahuja 
et al.\cite{0_Ahuja_2022} were added in stages along with any patterns that did 
not make it into Ahuja et al.'s original graph. This constituted version 1.1.

\textbf{Taxonomy Version 2:}
The second version of the taxonomy sought to reduce duplicates and condense 
the graph to a saturation point (no merge-able nodes exist). Though time 
constraints prevented us from reaching saturation, emergent communities were 
evident even after only a few iterations of the process described 
in Figure\ref{fig:ProcessFig}.

\textbf{Taxonomy Version 3:}
Although we are still discovering the occasional merge candidate, version 
3 of the graph is nearly saturated. This iteration of the taxonomy removed the
Author-type nodes to reduce the graph complexity and improve the ability of 
community detection to reveal prominent patterns and clarify their interrelation. 
It also eliminated single-node communities either by integration into larger 
communities, merging with identical nodes, or removal due to irrelevance. This 
is the version of the graph that we have based our analysis on.

\subsubsection{Node-Merging Process}
\begin{figure}[H]
    \centering
    \includegraphics[scale=0.35]{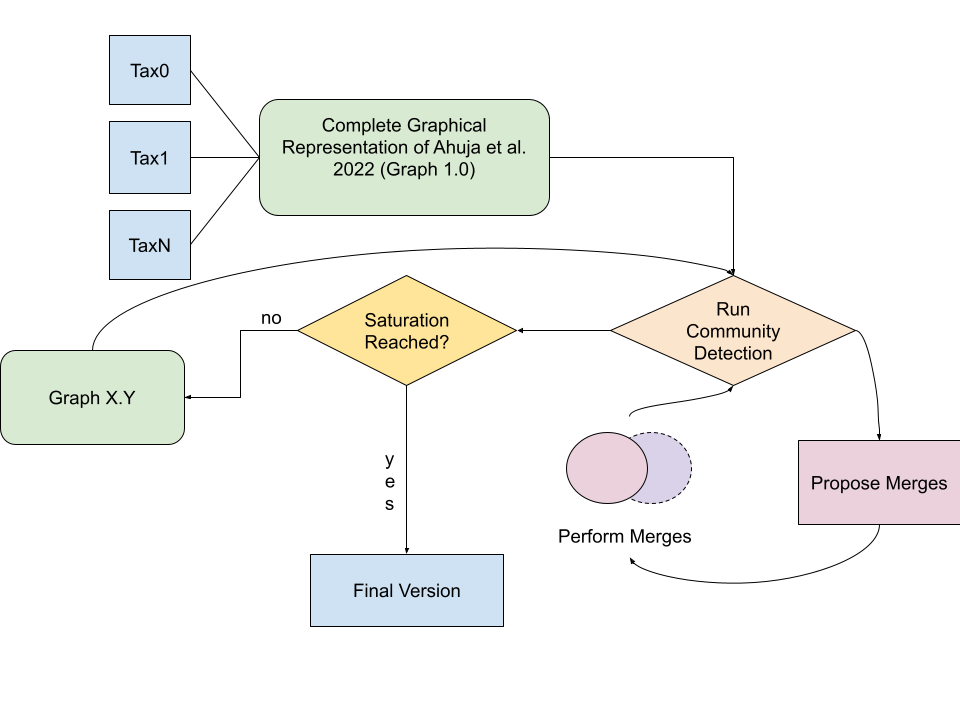}
    \caption{The construction and reduction process applied to 
    Ahuja et al.'s 2022 work.}
    \label{fig:ProcessFig}
\end{figure}

\begin{enumerate}[topsep=0pt]
	\item Utilizing Gephi, we ran the Fast Unfolding algorithm\cite{GEPHI} with 
a resolution of 1.0 on a randomized version of our graph to detect communities 
amongst the nodes. We noticed that running detection would often produce a varying 
number of communities so the detection was run five times with the number of 
communities discovered most often being selected at the end. Should there be a tie, 
the next tie breaking occurrence would be selected.
	\item After communities were detected, they were analyzed for patterns 
similar enough to have their nodes merged, or related enough to form an edge that 
was not previously identified in literature. When such nodes were found, 
they were flagged as a merge candidate and the process continued.
	\item Once all of the change candidates were identified, the reasoning 
behind their merge/new relation was explained, documented, and either approved or rejected.
	\item Approved merges and edges were enacted on the graph which would 
result in a new minor version and signify that the process could iterate once 
again.
	\item Once saturation of nodes is reached the graph has been condensed 
enough, and the iteration can cease.
\end{enumerate}
See Fig\ref{fig:ProcessFig} for a graphical representation of this process. 

\subsection{Detected Communities and Prominent Nodes}\label{Comms}
\begin{figure}
    \centering
    \includegraphics[scale=0.47]{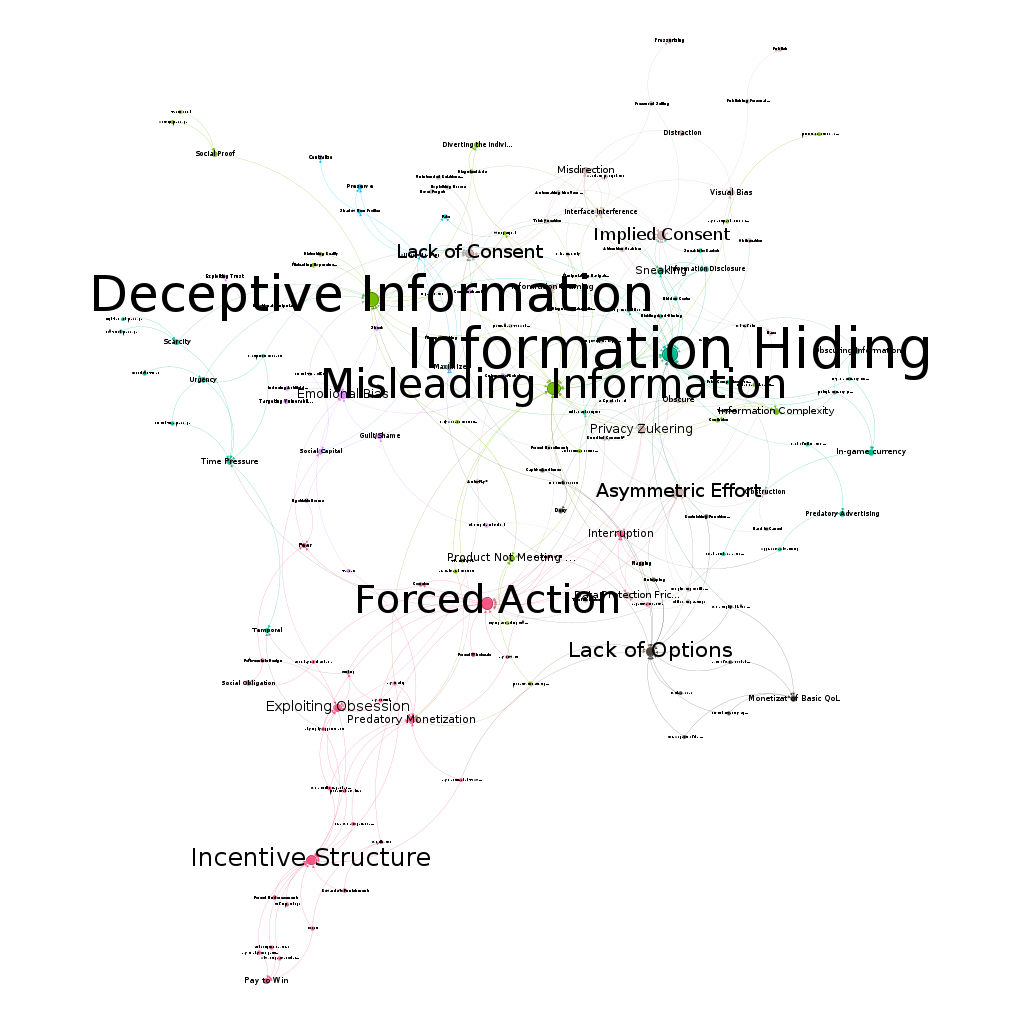}
    \caption{Full Integrated Taxonomy 3.2}
    \label{fig:FULL}
\end{figure}

The full integrated taxonomy is presented in Fig.\ref{fig:FULL}.In our research, we detected 10 communities centered around 4 major 
patterns (Information Hiding, Deceptive Information, Forced Action, and Misleading Information).
In this section we shall discuss these communities and their contained 
nodes in detail from largest to smallest community in terms of members. Due to space constraints the graphs representing in detail each community can be found in \href{https://docs.google.com/document/d/1eg65yB8xensTVLu3Rqzq1jrCMhmZjyh7BER7OHUXj9E/edit?usp=sharing}{Appendix available online}. 

\underline{\textbf{Forced Action} -
Main Pattern: Forced Action}
This community captures dark patterns which use their design to force or 
otherwise bend users into certain interaction patterns. Many of the DPs it 
contains come from the \textit{games} domain, through gamification however,
they exist in nearly any interactive system. Whether in the form of prompts 
that \textit{must} be forcefully dismissed by the user (Nagging\cite{6_Gray_2018}, 
Interruption\cite{2_Conti_2010}), monetary barriers to system use (Pay to Win/Skip \&
Pay or Wait/Grind\cite{14_Petrov_2021,3_Zagal_2013}), or barriers demanding data in 
exchange for the service(Impenetrable Wall\cite{10_CNIL_2020}), the patterns in 
this community restrict or coerce users into behaving a certain way.

\underline{\textbf{Deceptive \& Misleading Information} - Main Patterns: Deceptive}

\underline{Information, Misleading Information}
This community deals with cases where the information presented to the user 
is not necessarily \textit{false} but is deceptive or misleading in a way that 
meets the designers goals to influence the users. This includes cases where user 
interactions with the UI lead to unexpected or contradictory results (i.e. 
Wrong Signal\cite{10_CNIL_2020}, Manipulating Navigation\cite{2_Conti_2010}) and when 
the UI has lead the user to believe things about products that are untrue.

\underline{\textbf{Regarding Consent} -
Main Patterns: Lack of Consent, Implied Consent}

This community contains patterns which relate to users ability/inability to 
consent as well as scenarios where their consent was blatantly ignored.
Patterns like ``Automating the User Away"\cite{11_Gray_2020} remove the ability 
for users to consent while patterns like ``Trick Question"\cite{1_Brignull_2010} 
raise issue about whether consent was informed. This community also deals with 
the common scenario of default settings being skewed to meet the goals of 
the designers over the users through patterns like ``Bad Defaults"\cite{5_Bosch_2016},
``Default Sharing"\cite{10_CNIL_2020}. This community also seems to have captured 
cases where visual interference and bias have played a role in influencing 
user interaction as it contains patterns like Misdirection\cite{2_Conti_2010}, 
Interface Interference\cite{6_Gray_2018}, and Visual Bias\cite{0_Ahuja_2022}.

\underline{\textbf{Information Hiding} -
Main Pattern: Information Hiding}
With an in-degree of 27, Information Hiding is the most utilized pattern in the 
taxonomy. None of the other nodes in its community have near the same 
amount of utilizations though they appear to converge around the theme 
of \textit{monetization}, specifically in-game currencies. Seeing as 
in game currencies serve to hide, obscure, or otherwise disconnect user 
cognition of the \textit{actual} value of items for sale\cite{14_Petrov_2021}, this 
would make sense. Likewise, if we wanted to stop a user from comparing two 
products (Price Comparison Prevention\cite{1_Brignull_2010} or Comparison 
Obfuscation\cite{10_CNIL_2020}), hiding this information from the user will 
achieve this goal.

\underline{\textbf{Regarding Emotions} -
Main Pattern: Emotional Bias}
This community comprises patterns which induce and/or exploit user emotions. Emotional Bias is the engine behind patterns like ``Limited Time 
Offers"\cite{14_Petrov_2021} which can induce ``fear" of loss. ``Confirmshaming"
\cite{1_Brignull_2010} is another long-standing pattern in the DP space. It is 
difficult to determine when a line is crossed to make some of the patterns 
of this community exploitative. When we use an interface that suits our needs 
and helps us to achieve our goals, we feel good. Does this mean that the 
interface is utilizing dark patterns as it ``Induced Artificial Emotions"\cite{13_Mhaidli_2021}?
Not necessarily, good UI/UX design should by its nature be pleasant to interact 
with, or at least deliver an experience that both the user and the designer 
are expecting. An interface pattern containing an appeal for donation from the user if they enjoy the software would not be considered a dark pattern, even if it happens to evoke a feeling of shame in the user because the intention in the design is obvious and there is no hidden manipulation.  If emotions are induced such that they put the user in a state 
to the benefit of the designer, they cannot truly be said to be behaving as 
they would normally. Whether or not an offer is legitimately ``Limited Time", 
communicating this to the user with the goal of purposefully inducing fear/panic of missing an opportunity  
is emotionally manipulative and exploitative. 

\underline{\textbf{Lack of Options} -
Main Pattern: Lack of Options}
The ``Lack of Options"\cite{0_Ahuja_2022} community is quite similar to that of 
Forced Action. This is because when presented only a limited and curated set 
of potential interactions, users are essentially forced into choosing a path 
forward, none of which may be agreeable to them. A common example of this is 
in any ``Terms of service" agreement. This is not to say that all such 
agreements are innately dark, only that they can be used to achieve patterns like 
``Bundled Consent"\cite{12_Bongard_2021} where users must accept aspects of a 
system that they may not agree to in order to achieve others. This 
community also features many game-specific patterns to the effect of: 
``Pay or interact with an incomplete, under-powered, or useless system"\cite{14_Petrov_2021}.
This essentially infringes on user agency by offering the ultimatum: 
``Play by our rules, or do not play". Users could, of course, take their business 
elsewhere but this too runs into a ``lack of options" problem. Competitors 
may not deliver the desired quality and are likely offering their own ultimatum 
to users anyway, so again users are essentially forced into compliance for 
lack of desired options. 

\underline{\textbf{Regarding Privacy} - 
Main Pattern: Privacy Zukering}
This community captures patterns involved in tricking users into 
sharing as much data as possible.
Privacy Zukering is a pattern that has stood the test of time.  For example ``Obscure"\cite{5_Bosch_2016} and ``Information Framing"\cite{0_Ahuja_2022,7_NCC_2018} constitute patterns that can achieve
Privacy Zukering, whereas ``Forced Enrollment" \cite{9_Mathur_2019,5_Bosch_2016} 
and ``Safety Blackmail"\cite{10_CNIL_2020} are examples where it is applied. Though 
this community is small, it reveals the mechanics of data collection and forms 
that it might take.

\underline{\textbf{Pressurizing} - 
Main Patterns: Scarcity, Time Pressure, Urgency}
This community deals with patterns that pressurize the user. This is normally 
rooted in ``Time Pressure"\cite{0_Ahuja_2022} but is further specialized into 
instances of ``Scarcity" or ``Urgency"\cite{9_Mathur_2019}, both of which 
have denote actions with consequences if not acted upon \textit{soon}. Patterns in this community usually create an emotion - fear of missing an opportunity, so there are links between the patterns in these communities. 

\underline{\textbf{Friction} -
Main Pattern: Asymmetric Effort}
Previous version of the graph struggled to separate ``Asymmetric Effort"\cite{0_Ahuja_2022}
from ``Forced Action"\cite{0_Ahuja_2022,6_Gray_2018,7_NCC_2018,9_Mathur_2019}. 
Asymmetric effort simply deals with scenarios where the effort to preform an 
action are disproportionate to the effort required to undo them. Design friction 
describes ways in which effort can be increased/reduced within a UI, some of 
which are represented within this community (i.e. ``Obfuscating Settings"\cite{10_CNIL_2020},
Obstruction\cite{6_Gray_2018}.

\underline{\textbf{Bosch Outcasts} -
Main Patterns:  Maximize, Preserve}
Occasionally in the graph construction process, certain communities remain
around their defining author. This was often the case when the items defined 
in their taxonomy were not easily linked to rest of the graph. Such 
communities were labeled ``outcasts". We speculate that this community has 
remained outcast because of its unique construction. Reversing privacy-by-design 
principles to form their taxonomy members put them at a higher level of abstraction.
``Maximize"\cite{5_Bosch_2016} for example, is similar to the concept of ``Privacy
Zukering"\cite{1_Brignull_2010} but at a \textit{system} level, that is, a 
system can \textit{maximize} the amount of user data collected \textit{through}
Privacy Zukering. Similarly ``Preserve"\cite{5_Bosch_2016} exists on a system 
level and describes scenario's where user data is stored in a non-aggregated 
form\cite{5_Bosch_2016}. ``We never Forget``\cite{4_Greenberg_2014} and ``Unintended 
Relationships"\cite{4_Greenberg_2014} are examples of patterns (outside
of Bosch et al.'s) that may preserve data this way, though because the data 
lives on the server-side, this could not be discerned by a user. If anything, 
the fact that this community has remained outcast for 3 series of revisions 
suggests a gap in our reduction process.

\section{Discussion}
First we will revisit our research questions to
asses if our work was able to address them. Then we will look briefly at the 
implications of our work with respect to the construction of a Globally Harmonized 
System (GHS) for dark patterns formulated through heuristic evaluation before 
detailing our road-map for the project in the future.

\subsection{Research Questions Revisited}
Now that we have a better idea of what the integrated taxonomy contains, we will 
return to our research questions.
\newline

\textbf{RQ1}: What general Dark Pattern categories emerge from existing literature?

 The present of the graph (3.2), continues to reveal distinct DP communities as 
we outlined in section\ref{Comms}
\newline

\textbf{RQ2}: Can a process of heuristic evaluation be used to identify Dark 
Patterns based on emerging categories?

We attempted to form heuristics out of the resulting graph as early as version 
2.1 and though it did render results, they are only meant as a proof of concept
\ref{fig:POC}. Current work on this front is applying Straussian Grounded 
Theory\cite{GT_Strauss} using the Dark patterns in version 3.2 and their 
definitions as source material in an effort to contextualize the patterns and 
explicitly define their relation to one another. We will revisit this in the 
following section. 

\subsection{Future Work}
\begin{figure}[H]
    \centering
    \includegraphics[scale=0.25]{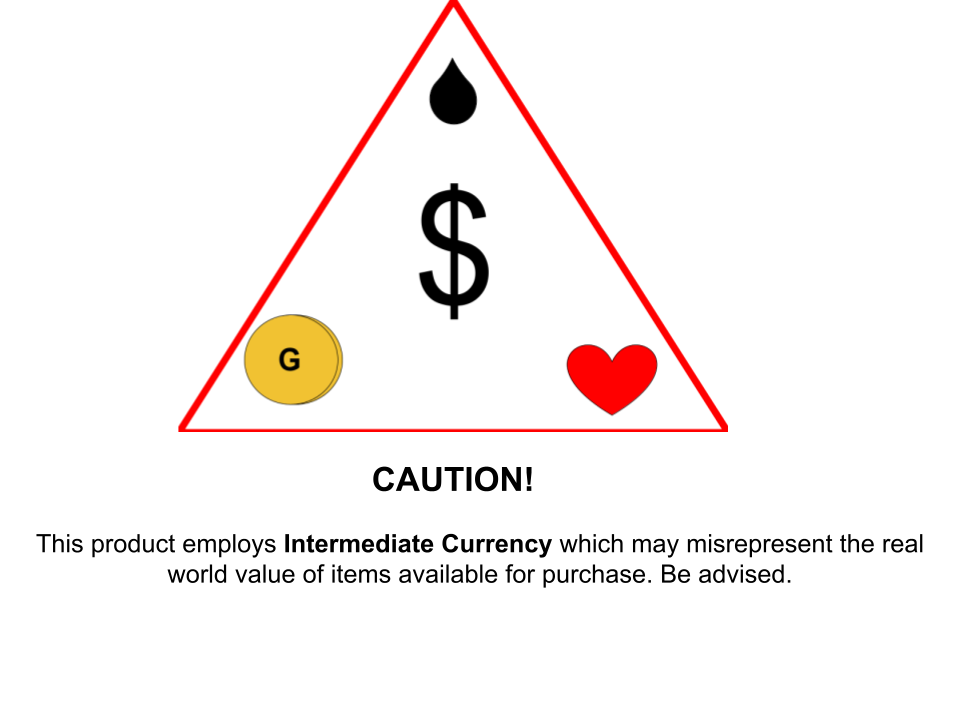}
    \caption{Example glyph: Intermediate Currency}
    \label{fig:POC}
\end{figure}

As we progress to the next stage of our work, applying Straussian Grounded 
Theory\cite{GT_Strauss} (SGT), we have discovered only a handful of new merge 
candidates. Because of this, we anticipate reaching a final set of Dark Patterns 
shortly after we complete the axial coding. Future work focuses on using SGT
to contextualize and define edge types (relationships) between dark pattern nodes
(i.e. consequence of, cause of, property of, etc.). Preliminary efforts 
on this front have already shown promising results. SGT has also helped in the identification and categorization of dark patterns, their properties, 
and dimensions.  Higher-level categories rendered by this process, along with the graph, will serve as a foundation to construct and \textit{justify} design heuristics capable of identifying malicious design. 
Finally, we will construct examples of a glyph-based hazard communication 
system by mapping heuristic violations to glyphs that can be included in the 
application/game description to inform users, similar to the example shown in Fig.\ref{fig:POC}. With these examples completed, we will  
evaluate the system's understandability, clarity, and applicability to malicious 
interfaces across different domains (e-commerce, social media, games) with law students, UI designers, and users. 
We hope that our work will inspire regulators, industry standards developers, 
software professionals, and UI designers to implement a GHS-similar standard 
notation to warn and protect users from dark patterns in interactive applications. 

\section{Conclusion}
Everyday interaction with computers has 
become an essential prerequisite to participating in our society.  If the 
majority of users are exposed to, but unaware of, dark patterns, the 
potential for harm and exploitation is unacceptably high. A domain-independent, 
universally understood language describing dark patterns is a critical 
prerequisite to creating regulation that can outlaw the most harmful dark patterns and also effective user education.

The integrated taxonomy presented here does not claim to be \textit{the ultimate} 
solution to the issue of hazard classification and communication in UI/UX design.
To define a language or classification system with that rigour would require an 
international, interdisciplinary team of experts akin to that which formed the 
GHS for hazardous materials. The contribution of our work is in demonstrating a method  
for creating an integrated 
taxonomy based on all available previous ones. By integrating the strongest examples of Dark Pattern taxonomies in a graph structure, bolstered with heuristics to  evaluate against, our work could serve as a starting point for developing a GHS for dark patterns. 

\section{Acknowledgements}
This research was supported by NSERC through the Discovery Grant Program RGPIN-2021-03521 to the second author.

\bibliographystyle{splncs04}
\bibliography{./2023_paper}

\end{document}